\documentstyle[12pt]{article}

\renewcommand{\theequation}{\thesection.\arabic{equation}}

\newcommand \beq{\begin{eqnarray}}
\newcommand \eeq{\end{eqnarray}}
\newcommand{\bea}{\begin{eqnarray*}}
\newcommand{\eea}{\end{eqnarray*}}
\newcommand{\be}{\begin{equation}}
\newcommand{\ee}{\end{equation}}
\newcommand{\ba}{\begin{array}}
\newcommand{\ea}{\end{array}}
\newcommand{\bc}{\begin{center}}
\newcommand{\ec}{\end{center}}

\begin{document}

\title{
The Poisson structure of the mean-field equations 
in the $ \Phi^4$ theory
}
\author{
C\'ecile Martin 
\\
\\
{
Groupe  de Physique Th\'eorique de l'Institut de Physique Nucl\'eaire  }\\
{  Institut de Physique Nucl\'eaire,} \\
{ F-91406 , Orsay Cedex, France  \footnote{E-mail address :
martinc@ipno.in2p3.fr}} 
}

\maketitle 
\vspace*{1cm}
\begin{abstract}
We show that the mean-field time dependent equations in the $ \Phi^4$
theory can be put into a classical non-canonical hamiltonian framework 
with a Poisson structure which is a generalization of the standard 
Poisson bracket. The Heisenberg invariant appears as a structural invariant
of the Poisson tensor. 
\end{abstract}

\date{}


\newpage

{\bc {\Large Poisson Structure in $ \Phi^4$ Theory} \ec}

\newpage

\section{ Introduction}

 The purpose of this paper is to give a formulation of the mean-field
 dynamical equations in a bosonic quantum field theory in terms of
 generalized Poisson brackets. The time-dependent Hartree-Bogoliubov (TDHB)
 equations are dynamical equations for the expectations values of one and
 two field operators. They have been used to study non-equilibrium evolution
 in $ \Phi^4$ theory \cite{JACKIW}. They are quantum dynamical
 equations. However, we will show that they can be put into a classical
 non-canonical hamiltonian framework with a Poisson structure which is a
 generalization of the standard Poisson bracket. 
 
 Non-canonical Poisson brackets have been found to be usefull in different
 contexts, including plasma and fluid dynamics \cite{WEINSTEIN} and
 condensed matter physics \cite{DV}. In relativistic field theories, it has
 been shown how to write the field equations in covariant Poisson bracket
 form \cite{MARSDEN}. In non-relativistic quantum theory, a Poisson
 structure has been recognized for the time-dependent mean-field dynamics of
 a system of fermions \cite{BV1} or a system of bosons \cite{BENAROUS}. A
 Poisson structure is especially useful to find structural invariants and
 to study stability properties. We will show how this approach can be
 developped in quantum field theory. 

 We consider the $ \Phi^4$ theory. In section 2, we introduce the
 dynamical variables of interest and we recall the TDHB equations which can
 be written in a compact form \cite{MARTIN}. We define the Poisson tensor in
 section 3 and we show the Poisson structure of the TDHB equations. Another
 definition of the Poisson tensor from a Lie algebra is considered in section
 4. In section 5, we recognize the Heisenberg invariant as a structural
 invariant of the Poisson tensor. Finally, in section 6, we write the
 linearized dynamical equations arround the HB static solution.

\section{The mean-field equations in the $ \Phi^4$ theory :}

\setcounter{equation}{0}

 We consider a self-interacting scalar field in a Minkowski metric described
 by the Hamiltonian $H=\int d^dx \: {\cal H}(\vec x)$ with
 \be \label{1.1}
 {\cal H}(\vec x)=\frac{1}{2}\Pi^2(\vec x)+\frac{1}{2}\left(\vec \nabla
 \Phi(\vec x) \right)^2 +\frac{m_0^2}{2}\Phi^2(\vec
 x)+\frac{b}{4}\Phi^4(\vec x) \ . \ee
 We work in $d$ spatial dimensions. The constants $m_0$ and $b$ are
 respectively the bare mass and the bare coupling constant. $\Phi(\vec x)$
 and $\Pi(\vec x)$ are the field operators in the Schr\"odinger
 representation. The initial state of the system is supposed to be given and
 equal to a statistical operator $D(t_0)$ which may represent a thermal
 equilibrium or a non-equilibrium situation. At zero temperature, it reduces
 to a projection operator.

  A time-dependent variational principle \cite{BV2} and the choice of gaussian
  operators for the trial space allow to derive time-dependent mean-field
  equations for the expectation values \cite{JACKIW,MARTIN,BV2}. A gaussian
  density-like operator ${\cal D}(t)$ is completly characterized by a vector
$\alpha(\vec x,t)$ and a matrix $\Xi(\vec x, \vec y,t)$ defined by : 
\be \label{1.2}
\alpha(\vec x,t)=\pmatrix{\varphi(\vec x,t) \cr -i \pi(\vec x,t) \cr} \ , \ee
 \be \label{1.3}
 \Xi(\vec x, \vec y,t)=\pmatrix{2G(\vec x,\vec y,t) & -i T(\vec x, \vec y,t)
 \cr -i T(\vec y, \vec x,t) & -2 S(\vec x,\vec y,t) \cr } \ , \ee 
 where 
\be \label{1.4} 
\varphi(\vec x,t)=\frac{1}{n(t)} \: Tr({\cal D}(t) \Phi(\vec x)) \ , \ee 
\be \label{1.5} 
\pi(\vec x,t) = \frac{1}{n(t)} \: Tr({\cal D}(t) \Pi(\vec x)) \ , \ee 
\be \label{1.6} 
G(\vec x,\vec y,t)= \frac{1}{n(t)} \: Tr({\cal D}(t) \bar  \Phi(\vec x) 
\bar \Phi(\vec y)) \ , \ee 
\be \label{1.7} 
S(\vec x,\vec y,t) = \frac{1}{n(t)} \: Tr({\cal D}(t) \bar \Pi(\vec x) 
\bar  \Pi(\vec y) ) \ , \ee 
\be \label{1.8} 
T(\vec x, \vec y,t) = \frac{1}{n(t)} \: Tr({\cal D}(t) 
( \bar \Phi(\vec x) \bar \Pi(\vec y) + 
\bar  \Pi(\vec y) \bar \Phi(\vec x) )) 
\ , \ee 
\be \label{1.9}
n(t)=Tr {\cal D}(t) \ , \ee
and, for any operator $O$ :
$\bar O = O-Tr({\cal D}(t) O)$.   

 The time-dependent Hartree-Bogoliubov (TDHB) equations for the $
 \Phi^4$ theory can be written in the following compact form \cite{MARTIN} : 
\be \label{1.10} 
i \: \dot \alpha = \tau \: w \ , \ee 
\be \label{1.11} 
i \: \dot \Xi = - \left[ \left( \Xi + \tau \right) 
\: {\cal H} \: 
\left( \Xi - \tau \right) - \left( \Xi - 
\tau \right) \: {\cal H} \: 
\left( \Xi + \tau \right) \right] \ . \ee
where $\tau$ is the $2 \times 2 $ matrix 
\be \label{1.12}
\tau = \pmatrix{0 & 1 \cr -1 & 0\cr } \ee
The vector $w$ and the matrix ${\cal H}$ are defined from the variation
of 
\be \label{1.12a}
<H>=\frac{1}{n(t)}Tr({\cal D}(t) H)=\int d^dx \: {\cal E}(\vec x,t) \ee
\be \label{1.13}
\delta <H> =\int d^d x \: \tilde w (\vec x,t)\: \delta \alpha(\vec x,t) - 
\frac{1}{2} \int d^dx d^dy \: tr \left({\cal H}(\vec x, \vec y,t) \delta 
\Xi(\vec y, \vec x, t) \right) \ . \ee
${\cal E}(\vec x,t)$ is the mean-field energy density. 
For the Hamiltonian (\ref{1.1}), we have : 
\be \label{1.14} 
w_1(\vec x,t) =\left( -\Delta + m_0^2 +\frac{b}{6} \: \varphi^2(\vec x,t) 
+ \frac{b}{2} \: G(\vec x, \vec x,t)\right) \: \varphi(\vec x, t)  \ee 
\be \label{1.15}
w_2(\vec x,t)=i \pi(\vec x,t)  \ee 
\be \label{1.15a} 
{\cal H}_{11}(\vec x,\vec y,t) = 
- \frac{1}{2}\left( - \Delta + m_0^2 + \frac{b}{2} \: 
\varphi^2(\vec x,t) + \frac{b}{2} \: G(\vec x,\vec x,t) \right) \: 
\delta(\vec x- \vec y)  \ee
\be \label{1.16}
{\cal H}_{22}(\vec x, \vec y,t)=\frac{1}{2}\delta(\vec x-\vec y) \ee
\be 
{\cal H}_{12}(\vec x, \vec y,t)=0 \quad \ , \quad 
{\cal H}_{21}(\vec x, \vec y,t)=0 \ee

The static HB solution is caracterized by  $ w_0=0$, i. e. in the uniform
case 
\be \label{1.17}
(m_0^2+\frac{b}{6}\varphi_0+\frac{b}{2}G_0(x,x))\varphi_0=0 \ee
and the gap equation 
\be \label{1.18}
 \Xi_0=-\tau \: \coth(\beta  {\cal H}_0 \tau) \ee
which can also be written as 
\be \label{1.19}
\beta  {\cal H}_0=\frac{1}{2}\: \tau \ln \left(\frac{ \Xi_0+\tau}
{ \Xi_0-\tau}\right) \ . \ee
More explicitly, in the uniform case, by working  in the momentum space,   
\be \label{1.20}
 \Xi_0(p)=\pmatrix{\frac{1}{\omega(p)} \: \coth \frac{\beta \omega(p)}{2}
& 0 \cr 0 & -\omega(p) \: \coth \frac{\beta \omega(p)}{2} \cr } \ , \ee
with
$\omega^2(p)= p^2+m_0^2+\frac{b}{2}\varphi_0^2 +
\frac{b}{2}  G_0(\vec x, \vec x) $.  

In the following we will show that the mean-field equations (\ref{1.10}) and
(\ref{1.11}) have a Poisson structure. 

\section{The Poisson tensor and the Poisson structure of the mean-field
equations} 

\setcounter{equation}{0}

 We consider first the uniform case. A more general construction of the
 Poisson tensor in the non-uniform case will be given in the next section.
 We gather in a single vector $z^{\mu}, \mu=1,2,3$
 the expectation values $<1>, \alpha_j$ and
$  \Xi_{ij}(p)$, where $\Xi_{ij}(p)$ is the Fourrier transformed of $\Xi(\vec
x, \vec y,t)$  : 
\be 
\Xi_{ij}(\vec x, \vec y)=\int \frac{d^dp}{(2 \pi)^d}\: e^{i \vec p . (\vec y-
\vec x)} \: \Xi_{ij}(p) \ . \ee
We define the Poisson tensor  ${\cal C}^{\mu
  \nu} \equiv \{ z^{\mu},z^{ \nu} \} $ in the following way :
  \be \label{3.1} \ba{llllllll}
  & {\displaystyle
  {\cal C}^{22}_{ij}=\tau_{ij} \quad \ , \quad
  {\cal C}^{23}_{i,jk}=0 } \\
  & {\displaystyle
  {\cal C}^{33}_{11,12}(p)={\cal C}^{33}_{11,21}(p)=4 \: \Xi_{11}(p) } \\
  & {\displaystyle
  {\cal C}^{33}_{12,11}(p)={\cal C}^{33}_{21,11}(p)=-4 \:\Xi_{11}(p) } \\
  & {\displaystyle   {\cal C}^{33}_{11,22}(p)=8 \: \Xi_{12}(p) } \\
  & {\displaystyle   {\cal C}^{33}_{22,11}(p)=-8 \: \Xi_{12}(p) } \\
  & {\displaystyle
  {\cal C}^{33}_{12,22}(p)={\cal C}^{33}_{21,22}(p)=4 \: \Xi_{22}(p)
  } \\
  & {\displaystyle 
  {\cal C}^{33}_{22,12}(p)={\cal C}^{33}_{22,21}(p)=-4 \: \Xi_{22}(p) } \\
 & {\displaystyle
  {\cal C}^{33}_{12,12}(p)={\cal C}^{33}_{12,21}(p)=0 } \ea \  \ee
  The other matrix elements of this tensor are equal to zero.
${\cal C}^{22}$ is a constant in the momentum space, while
${\cal C}^{33}$ depends on the momentum   $p$. This Poisson tensor is a
linear function of the dynamical variables $\Xi_{ij}(p)$. Let us note that it
does not depend on the value of the condensat $\alpha$. 

The Poisson structure associated to the Poisson tensor ${\cal C}^{\mu \nu}$
is defined according to
 \be \label{3.2}
   \{f,g\} \equiv \frac{1}{i} \: \frac{\partial f}{\partial z^{\mu}} \:
   {\cal C}^{\mu \nu} \: \frac{\partial g}{\partial z^{\nu}} \ , \ee
   for any functions $f$ and $g$ of the dynamical variables $z^{\mu}$. 

We check that the operation $\{f, g \}=k$ which associates to two
   functions $f$ and $g$ a third function $k$ is bilinear and antisymmetric.
   It satisfies also Jacobi's identity and Leibnitz's derivation rule 
   \cite{DIRAC}.
   (When the Poisson structure is not defined from a Lie algebra, we have to
   check these properties.) 

In particular, we have
  \be \label{3.3}
  \{\alpha_i,g\}  =  \frac{1}{i}\: {\cal C}^{22}_{ij}\:
  \frac{\partial g}{\partial \alpha_j}
   =  \frac{1}{i}\: \tau_{ij}
  \frac{\partial g}{\partial \alpha_j}  \ . \ee
From the definition $w_i\equiv \frac{\partial {\cal E}}{\partial \alpha_i}$, 
we can write the first TDHB equation (\ref{1.10}) as : 
\be \label{3.4} 
\dot \alpha_i=\{ \alpha_i, {\cal E} \} \ . \ee 

We have also :
 \be \label{3.5}
 \{ \Xi_{ij},g\}= \frac{1}{i}\: {\cal C}^{33}_{ij,kl}\:
 \frac{\partial g}{\partial \Xi_{kl}} \ , \ee
If we choose $g={\cal E}$, with the definition (\ref{3.1}) of ${\cal C}^{33}$
and the definition (\ref{1.13}) of the matrix ${\cal H}$, we obtain :  
 \be \label{3.6} \ba{lll}
 & {\displaystyle 
 i \: \{\Xi_{11},{\cal E} \}=-4 \: \Xi_{12}(p) \: {\cal H}_{22}(p) } \\
 & {\displaystyle 
 i \: \{\Xi_{12},{\cal E} \}=2 \: \Xi_{11}(p) \: {\cal H}_{11}(p) 
 -2 \:  \Xi_{22}(p)
 {\cal H}_{22}(p) } \\
 & {\displaystyle 
 i \: \{\Xi_{22},{\cal E} \}=4 \: \Xi_{12}(p) \: {\cal H}_{11}(p) } \ea \ . \ee
 The second TDHB  equation (\ref{1.11}) can be put in the form :  
 \be \label{3.7} 
 \dot \Xi_{ij}=\{\Xi_{ij}, {\cal E}\} 
 \ee 

 The TDHB equations in the $ \Phi^4$ theory can therefore be written
 as classical dynamical equations, the mean-field 
 energy density ${\cal E}(\alpha,
 \Xi)$ playing the role of a classical Hamiltonian with the non-symplectic
 Poisson structure defined by (\ref{3.1}). 

 It is actually more useful to consider the free-energy density 
 \be \label{3.8}
 {\cal F}(\alpha, \Xi) ={\cal E}(\alpha, \Xi)-\beta^{-1}{\cal S}(\Xi) \ee
 as a classical Hamiltonian instead of ${\cal E}$. The von-Neuman entropy
 density ${\cal S}$ is defined by : 
 \be \label{3.8a}
 {\cal S} \equiv 
 -\frac{Tr {\cal D} \log {\cal D}}{Tr {\cal D}} +\log Tr {\cal D} \
 . \ee
 For a gaussian density matrix it can be written as a
 function of the matrix $\Xi$ as : 
\be \label{3.9} \ba{ll}
{\cal S}=\int \frac{d^3p}{(2 \pi)^3}  
\: & {\displaystyle \left[
\frac{1}{4}\: tr\left\{ \left(\frac{1}{2}\: (1-\tau)\: \Xi
\: (1-\tau) -1 \right) \: \log \left(\frac{\Xi+\tau}{\Xi-\tau}\right)
\right\} \right. } \\ 
& {\displaystyle \left. + \frac{1}{2} \: \log \det \left\{ \frac{1}{4}\: 
\pmatrix{ 1& 1 \cr 1 & -1\cr } \: \Xi \: (1-\tau) -\frac{1}{2} \: 
\pmatrix{0&1\cr 1 & 0\cr } \right\} \right] }  \ea \ , \ee
or
\be \label{3.10}
{\cal S}=\int \frac{d^3p}{(2 \pi)^3} \: \left[ 
\frac{1}{2}\:\sqrt I \log (\frac{\sqrt I+1}{\sqrt I-1})   
+ \frac{1}{2} \: \log (I-1)  \right]   \ , \ee
where I is the Heisenberg invariant  
\be \label{3.11} \ba{ll}
I(x,y)=\int d^3z \: & {\displaystyle \left( 4 <\bar \Phi(x) \bar \Phi(z)> 
<\bar \Pi(z) \bar \Pi(y)> \right. } \\
& {\displaystyle \left. -<\bar \Phi(x) \bar \Pi(z)+ \bar \Pi(z) \bar \Phi(x)>
\: <\bar \Phi(z) \bar \Pi(y) + \bar \Pi(y) \bar \Phi(z)> \right) } \ea \ , \ee 
\be \label{3.12} 
I(x,y)=\int d^3z \: \left(-\Xi_{11}(x,z)\: \Xi_{22}(z,y) + \Xi_{12}(x,z) \: 
\Xi_{12}(z,y) \right) \ . \ee
For a pure state, we have : $
\Xi_{11} \: \Xi_{22}- \Xi_{12}^2+1=0 $ or  $
I(x,y)=\delta(x-y) $. 

We have $
  \{ {\cal S}( \Xi),\alpha\}=0$ and  $ \{ {\cal S}( \Xi),\Xi\}=0 $.
  Therefore, the TDHB equations (\ref{3.4}) and (\ref{3.7}) can be written
  as :
  \be \label{3.13}
   \dot \alpha_i=\{ \alpha_i, {\cal F}(\alpha, \Xi) \} \ , \ee
  \be \label{3.14}
  \dot \Xi_{ij}=\{\Xi_{ij}, {\cal F}(\alpha, \Xi) \} \ .
 \ee
 The free-energy density ${\cal F}(\alpha, \Xi)$ is a constant of the motion
 like ${\cal E}(\alpha, \Xi)$ and ${\cal S}(\Xi)$.  
  The static solution $\alpha_0, \Xi_0$ (see eq. (\ref{1.17}) and
  (\ref{1.18}))
  of the TDHB equations is a minimum of ${\cal F}$ (at zero temperature, it
  is a minimum of ${\cal E}$). In section 6, we will derive the linearized
  mean-field equations arround the static solution. 
  
   
\section{D\'efinition of the Poisson tensor from a Lie algebra  } 

\setcounter{equation}{0}

 In this section we  consider the non uniforme case and we work with the
 space indices. We define the following operators 
 ${\cal O}^{\mu}$  : 
\be \label{4.1} 
{\cal O}^1=1 \ee 
\be \label{4.2}
{\cal O}^2(\vec x)=\pmatrix{\Phi(\vec x) \cr -i \Pi(\vec x) \cr } \ee 
\be \label{4.3}
{\cal O}^3(\vec x, \vec y)=\pmatrix{2 \: \bar \Phi(\vec x) \bar \Phi(\vec y) & 
-i(\bar \Phi(\vec x) \bar \Pi(\vec y)+\bar \Pi(\vec y) \bar \Phi(\vec x) )
\cr -i(\bar \Phi(\vec y) \bar \Pi(\vec x)+\bar \Pi(\vec x) \bar \Phi(\vec y) )
& -2 \: \bar \Pi(\vec x) \Pi(\vec y) \cr} \ee
where $\bar {\cal O}={\cal O}-<{\cal O}>$ for any operator ${\cal O}$. 

It is easy to show that these operators satisfy a Lie algebra : 
\be \label{4.4}
[{\cal O}^{\mu},{\cal O}^{\nu}]=f^{\mu \nu}_{\sigma} \: {\cal O}^{\sigma} \ee 
More explicitly : 
\be \label{4.5}
[{\cal O}^2_i(\vec x),{\cal O}^2_j(\vec y)]=(f^{22}_1)_{ij}(\vec x, \vec y)
\: {\cal O}^1 \ee 
\be \label{4.6}
[{\cal O}^2_i(\vec x),{\cal O}^3_{jk}(\vec y, \vec z)]=\int d^du \: \sum_m
(f_2^{23})^m_{i,jk}(\vec x,\vec y,\vec z;\vec u) \: {\cal O}_m^2(\vec u) + 
(f^{23}_1)_{i,jk}(\vec x, \vec y, \vec z) \: {\cal O}^1 \ee 
\be \label{4.7}
[{\cal O}^3_{ij}(\vec x, \vec y), {\cal O}_{kl}^3(\vec z, \vec t)]=
\int d^du \: d^dv \: 
\sum_{m,n} (f_3^{33})^{mn}_{ij,kl}(\vec x, \vec y, \vec z, \vec t;\vec u,
\vec v) \: {\cal O}^3_{mn}(\vec u, \vec v) \ee
The structure constants $f^{22}_1,f^{23}_2,f^{33}_3$ are delta functions in
position space while 
 $f^{23}_1$ depends also on the expectation values 
 $<{\cal O}_i^2(\vec x)>$. The other structure constants 
$f^{\mu \nu}_{\sigma}$ are equal to zero. For instance, we have : 
\be \label{4.7a}
f_1^{22}(\vec x, \vec y)=\tau \: \delta(\vec x - \vec y) \quad \ , \quad 
(f_1^{23})_{1,12}(\vec x, \vec y, \vec z)=-2 \varphi(\vec x) \:
\delta(\vec x -\vec y) \delta(\vec x -\vec z) \ . \ee

We then consider the expectation values $z^{\mu}\equiv <{\cal O}^{\mu}>$
and we define the Poisson bracket ${\cal C}^{\mu \nu}\equiv
\{ z^{\mu},z^{\nu}\}$ from the Lie algebra characterized by the structure
constant $f$ : 
\be \label{4.8}
{\cal C}^{\mu \nu}\equiv
\{ z^{\mu},z^{\nu}\}=f^{\mu \nu}_{\sigma} z^{\sigma} \ee
Similarly to the structure constant $f$, the tensor 
${\cal C}^{\mu \nu}$ has space indices. 

In the uniform case, by working in the 
momentum space, we can show that the only
non-vanishing commutators between the operators ${\cal O}^{\mu}$ 
can be written as : 
\be \label{4.9}
[{\cal O}_i^2,{\cal O}_j^2]=\tau_{ij} 
\ee
\be \label{4.10}
[{\cal O}^2_i,{\cal O}^3_{jk}(\vec p=0)]=\sum_m (f_2^{23})^m_{i,jk} \: 
{\cal O}^2_m +(f^{23}_1)_{i,jk} \: {\cal O}^1 \ee
\be \label{4.11}
[{\cal O}^3_{ij}(\vec p),{\cal O}^3_{kl}(\vec p)]= 
\sum_{mn} (f^{33}_3)^{mn}_{ij,kl} 
\: {\cal O}^3_{mn}(\vec p) \ee
where the structure constants $f$ do not depend on the momentum $\vec
p$. In this way we find again the Poisson tensor defined in section 3. 

\section{Heisenberg Invariant }

\setcounter{equation}{0}

Another presentation of the Poisson tensor can be useful to find the
structural invariants . We still consider the uniform case and we work in
the momentum space.  By chosing the coordinates 
$<\Phi>,<\Pi>,<\bar \Phi \bar \Phi>(p),
 <\bar \Pi \bar \Pi>(p),
 <\bar \Phi \bar \Pi+\bar \Pi \bar \Phi>(p)$, 
 the Poisson tensor can be written as  a $5 \times 5$ matrix, build from two
 blocks : the symplectic matrix 
$\pmatrix{0&1\cr -1 & 0 \cr}$ and the  $3 \times 3$ antisymmetric matrix 
 defined by the elements : 
 \be \label{5.1}
 \tilde {\cal C}^{33}_{11,12}(p)\equiv \{<\bar \Phi \bar \Phi>(p), 
 <\bar \Phi \bar \Pi+\bar \Pi \bar \Phi> (p)\}=
 4  \: <\bar \Phi \bar \Phi>(p) \ee 
\be \label{5.2} 
 \tilde 
 {\cal C}^{33}_{12,22}(p)\equiv \{ <\bar \Phi \bar \Pi+\bar \Pi \bar \Phi>(p), 
 <\bar \Pi \bar \Pi>(p) \}=
 4  \: <\bar \Pi \bar \Pi>(p) \ee 
\be \label{5.3} 
 \tilde {\cal C}^{33}_{11,22}(p)\equiv \{ <\bar \Phi \bar \Phi>(p), 
 <\bar \Pi \bar \Pi>(p) \}=
 2  \: <\bar \Phi \bar \Pi + \bar \Pi \bar \Phi>(p) \ , \ee
 that is 
 \be \label{5.4}
 M=\pmatrix{0&2 \: <\bar \Phi \bar \Pi+\bar \Pi \bar \Phi>(p)&
 4  \: <\bar \Phi \bar \Phi>(p) \cr
 -2  \: <\bar \Phi \bar \Pi+\bar \Pi \bar \Phi>(p)
 & 0 & -4  \: < \bar \Pi \bar \Pi>(p) \cr
 -4  \: < \bar \Phi \bar \Phi>(p) & 4  \: <\bar \Pi \bar \Pi>(p) &  0 \cr} 
 \ee

 The rank of the Poisson tensor is a function of the dynamical variables.
 Near a regular point, the rank of the Poisson tensor is constant. 
We can find coordinates (Q,P,C) such as the matrix  $M$ can
 put in the form :
 \be \label{5.5}
 M'=\pmatrix{0&1&0 \cr -1 &0&0 \cr 0&0&0\cr } \ . \ee
 The Poisson brackets in this coordinates system are given by the
 symplectic form
 \be \label{5.6}
 \{f,g\}=\frac{\partial f}{\partial Q} \: \frac{\partial g}{\partial P}
 -\frac{\partial f}{\partial P}\: \frac{\partial g}{\partial Q} \ . \ee
 We have therefore  {\it locally } a canonical Hamiltonian system which
 depends on one parameter C. The function of the dynamical variables which
 is equal to C is a Casimir function: it satisfies
 $\{f, C\}=0$ for any function $f$ of the dynamical variables 
 \cite{WEINSTEIN}.  
 
 The coordinate C corresponding to the eigenvalue 0 of the Poisson tensor
 defined by the matrix  (\ref{5.4}) is : 
 \be \label{5.7} 
 C(p)=   4 \: <\bar \Phi \bar \Phi>(p) \: <\bar \Pi\bar \Pi>(p)  
 -<\bar \Phi \bar \Pi+\bar \Pi \bar \Phi>(p) \:  
 <\bar \Phi \bar \Pi+ \bar \Pi \bar \Phi>(p)   \ . \ee 
 We recognize the Heisenberg invariant (\ref{3.11}) and we have : 
 \be \label{5.8}
 \dot C(p)=0 \ee

\section{Linearized equations}

\setcounter{equation}{0}

 At the HB solution  $z^{\mu}_0$
 which is a regular point of the Poisson tensor, the matrix defined by :
   \be \label{6.1}
   {\cal B}^{\mu \nu} \equiv
   \frac{\partial^2 {\cal F}}{\partial z^{\mu} \partial z^{\nu}}
   \Big \vert_{z_0} \ee
   is positive definite. The free-energy density ${\cal F}$ being conserved
   by the flow defined with the Poisson brackets (\ref{3.2}), this implies
   that $z^{\mu}_0$ is a stable equilibrium point.  
 
 The TDHB equations write : 
\be \label{6.2}
\dot \alpha_i =\{\alpha_i, {\cal F}\}=\frac{1}{i}\: {\cal
C}^{22}_{ij}(\Xi) \: \frac{\partial {\cal F}}{\partial \alpha_j} \ee
\be \label{6.3}
\dot \Xi_{ij}=\{\Xi_{ij},{\cal F}\}=\frac{1}{i}\: {\cal C}^{33}_{ij,kl}
(\Xi) \: \frac{\partial {\cal F} }{\partial \Xi_{kl}} \ee
where we have explicitly written the dependence of the Poisson tensor on the
matrix $\Xi$. Let us develop these equations arround the static HB solution
$\alpha_0, \Xi_0$ at the first order in
$\delta \alpha=\alpha-\alpha_0$  and $ \delta \Xi=\Xi-\Xi_0$. 
Because $\alpha_0,
\Xi_0$ is a minimum of ${\cal F}$, the variation of  ${\cal C}$ arround 
$\alpha_0, \Xi_0$ does not contribute. By introducing  
\be \label{6.4}
{\cal F}^{(2)}(\alpha, \Xi)={\cal F}^{(2)}(\alpha_0, \Xi_0) + 
\frac{1}{2} \: \delta z^{\mu} \delta z^{\nu} \: \frac{\partial^2 {\cal F}}
{\partial z^{\mu} \partial z^{\nu}} \big \vert_{z_0} \ , \ee
we obtain : 
\be \label{6.5} 
\frac{d \delta \alpha_i}{dt}=\{\delta \alpha_i, {\cal F}^{(2)}(\alpha, \Xi) 
\}_{0}=\frac{1}{i}\:  {\cal
C}^{22}_{ij}(\Xi_0) \: \frac{\partial {\cal F}^{(2)}}
{\partial (\delta \alpha_j)} \ee
\be \label{6.6} 
\frac{d \delta \Xi_{ij}}{dt}(q)=\{\delta \Xi_{ij},
{\cal F}^{(2)}(\alpha, \Xi)\}_{0} =\frac{1}{i}\: {\cal C}^{33}_{ij,kl}
(\Xi_0)(q) \: \frac{\partial {\cal F}^{(2)} }{\partial
(\delta \Xi_{kl})} \ee
where the indice $0$ for  the Poisson bracket indicates that the Poisson
tensor ${\cal C}$ has to be evaluated at the point $\Xi_0$. In this way, we
obtain the small oscillations equations arround the static HB solution also
called RPA equations. These equations are of Hamilton type with the free
energy developped to second order
${\cal F}^{(2)}(\alpha, \Xi)$ playing the role of the hamiltonian and with
${\cal C}(\Xi_0)$ as Poisson tensor.  ${\cal F}^{(2)}$ is a constant of the
linearized equations of motion and has a local minimum at $\alpha_0, \Xi_0$.
Therefore the linearized dynamical equations at $z^{\mu}_0$ have $0$ as
stable equilibrium. 

The small oscillations equations (\ref{6.5}) and (\ref{6.6}) can be written
as : 
\be \label{6.7}
i \frac{d \delta \alpha_i}{dt}={\cal C}^{22}_{ij}(\Xi_0)\left( 
{\cal B}^{22}_{jk} \delta \alpha_k + \: \int \frac{d^3q}{(2
\pi)^3} \: {\cal B}^{23}_{j,kl}(q) \: \delta \Xi_{lk}(q) \right) \ee 
\be \label{6.8}
i \frac{d \delta \Xi_{ij}(q)}{dt}={\cal C}^{33}_{ij,kl}(\Xi_0)(q) \: 
\left (  \: {\cal B}^{32}_{lk,m}(q)\: \delta \alpha_m + 
\int \frac{d^3k}{(2\pi)^3} \: {\cal B}^{33}_{lk,mn}(q,k) \: \delta 
\Xi_{nm}(k) \right) \ee
The expression of the matrix elements of ${\cal B}^{\mu \nu}$ in the
$ \Phi^4$ theory is given in appendix.  
In equations (\ref{6.7}) and (\ref{6.8}), we have considered small
oscillations arround the uniform static HB solution : 
\be \label{6.9}
 \alpha(\vec p,t)=\alpha_0 \delta(\vec p) +\delta \alpha(\vec p,t) \ee
\be \label{6.10}
\Xi(\vec p_1,\vec p_2,t)=\Xi_0(\vec  p_1) \delta(\vec p_1+ \vec p_2) + 
\delta \Xi(\vec p_1,
\vec p_2,t) \ . \ee
where 
\be \label{6.11}
\Xi(\vec p_1,\vec p_2,t)=
\int d^3x \: d^3y \: e^{-i\vec p_1 \cdot \vec x-i \vec  p_2 \cdot \vec y}
\: \Xi(\vec x,\vec y,t) \ . \ee
We have then introduced the total momentum
$\vec P=\vec p_1+\vec p_2$ and the relative momentum
$\vec q=\frac{1}{2}(\vec p_1-\vec p_2)$ and we have written the equations for
a total momentum $\vec P=\vec 0$. 

Equations (\ref{6.5}) and (\ref{6.6}) could also have been obtained by
linearizing the original expressions (\ref{1.10}) and (\ref{1.11}) arround the
static solution :  
\be \label{6.12}
i \delta \dot \alpha =\tau \delta w \ee 
\be \label{6.13}
i \delta \dot \Xi =2 (\delta \Xi \: {\cal H} \: \tau -\tau \: {\cal H} \: 
\delta \Xi) +
2 ( \Xi_0 \: \delta {\cal H} \: \tau -\tau \: \delta {\cal H} \: \Xi_0) \ . \ee
We define the matrices $t, T, r, R$ (which are the analogs of the RPA matrix
${\cal R}$ of reference \cite{BV2}) by expanding $<H>$ up to the second order, or
the vector $w$ and the matrix ${\cal H}$ up to the first order : 
\be \label{6.14}
\delta w_i(\vec x,t)=\int d^d y \: t_{ij}(\vec x, \vec y,t) \: \delta
\alpha_j(\vec y,t) - \frac{1}{2} \int d^dy d^dz \: T_{i,jk}(\vec x, \vec y,
\vec z,t) \: \delta \Xi(\vec z, \vec y,t) \ee 
\be \label{6.15}
\delta {\cal H}_{ij}(\vec x, \vec y,t)=\int d^d z \: r_{ij,k}(\vec x, \vec
y,\vec z,t) \delta \alpha_k(\vec z,t) -\frac{1}{2} \int d^dz d^du \:
R_{ij,kl}(\vec x, \vec y, \vec z, \vec u, t) \: \delta \Xi_{lk}(\vec u, \vec
z, t) \ . \ee
We therefore obtain a relation between these matrices
and the matrix ${\cal B}$ of the second derivatives of the free energy at
the minimum. This relation involves the Poisson tensor evaluated at $\Xi_0$. 

From the equations (\ref{6.7}) and (\ref{6.8}), we can study the existence of
bound states at finite temperature both in the symmetric phase and in the
broken phase. This will be a generalization to finite temperature of the
results obtained by Kerman and Lin at zero temperature \cite{KERMAN}.

To conclude, we have shown that the time-dependent mean-field equations in
$ \Phi^4$ theory can be written as classical dynamical equations with
a non-symplectic Poisson structure, the free-energy density playing the role
of a classical Hamiltonian. The Heisenberg invariant appears as a structural
invariant of the Poisson structure. This Poisson structure can be useful to
discuss stability properties of trajectories characterized by the condensat
and the fluctuations of the quantum field at finite temperature. The
occurrence of a vanishing eigenvalue for the matrix ${\cal B}$ at a critical
temperature for a given coupling constant is the signal of a phase
transition from the phase with broken symmetry to the symmetric phase. 

\vspace*{1cm} 

{\bf Aknowledgements}

I wish to thank Marcel V\'en\'eroni for very helpful conversations. 
\vspace*{1cm}

{\bf Appendix}

\setcounter{equation}{0} 
\renewcommand{\theequation}{A.\arabic{equation}}

 In this appendix we give the expression of the elements of the matrix
 ${\cal B}$ in the case of the $ \Phi^4$ theory. 
\be \label{A.1}
{\cal B}^{22}_{11}=m_0^2+\frac{b}{2}\: \varphi^2_0 + \frac{b}{2} \: 
G_0(\vec x, \vec x) \quad \ , \quad 
{\cal B}^{22}_{22}=-1 \quad \ , \quad 
{\cal B}^{22}_{12}={\cal B}^{22}_{21}=0 \ee
\be \label{A.2}
{\cal B}^{23}_{1,11}=\frac{\partial^2 {\cal E}}{\partial \alpha_1 
\partial \Xi_{11}(p)}=\frac{b}{4}\: \varphi_0 \ee 
\be \label{A.3}
{\cal B}^{33}_{ij,kl}(p,q)=\frac{\partial^2 {\cal E}}{\partial \Xi_{ji}(p) 
\partial \Xi_{lk}(q)}-\frac{1}{\beta} \: \frac{\partial^2 {\cal S}}
{\partial \Xi_{ji}(p) \partial \Xi_{lk}(q)} \: \delta^3(p-q) \ee 
\be \label{A.4}
{\cal B}^{33}_{11,11}(p,q)=\frac{b}{16}+\frac{1}{4 \beta}\: 
\omega^2(p)\: (\frac{x}{\coth x}+\frac{1}{\coth^2x-1}) \delta^3(p-q) \ee 
\be \label{A.5}
{\cal B}^{33}_{11,22}(p,q)=\frac{1}{4 \beta} 
\: (\frac{x}{\coth x}-\frac{1}{\coth^2x-1}) \delta^3(p-q) \ee
\be \label{A.6}
{\cal B}^{33}_{22,22}(p,q)=\frac{1}{4 \beta \omega^2(p)} 
\: (\frac{x}{\coth x}+\frac{1}{\coth^2x-1}) \delta^3(p-q) \ee
\be \label{A.7}
{\cal B}^{33}_{12,12}(p,q)=-\frac{1}{\beta}\: \frac{x}{\coth x} \: 
\delta^3(p-q) \ee
where $x=\frac{\beta \omega(p)}{2}$ and  
\be \label{A.8} 
\omega^2(p)=p^2+m_0^2+\frac{b}{2} \varphi_0^2 + \frac{b}{2} G_0(\vec x, \vec
x, t) \ee
The other elements of the matrix  ${\cal B}$ are equal to zero.

\end{document}